\documentclass[manuscript]{aastex62}
\usepackage{natbib}
\begin{document}

\title{A Very Compact Extremely High Velocity Flow toward MMS 5 / OMC-3 Revealed with ALMA}

\author{Yuko Matsushita$^1$, Satoko Takahashi$^{2, 3, 4}$, Masahiro N. Machida$^1$ and Kohji Tomisaka$^{4, 5}$}
\affil{$^1$ Department of Earth and Planetary Sciences, Faculty of Sciences, Kyushu University, Fukuoka 819-0395, Japan; yuko.matsushita.272@s.kyushu-u.ac.jp\\
$^{2}$ Joint ALMA Observatory, Alonso de C{\'{o}}rdova 3107, Vitacura, Santiago, Chile\\  
$^{3}$ NAOJ Chile Observatory, Alonso de C{\'{o}}rdova 3788,Oficina 61B,Vitacura, Santiago, Chile\\ 
$^{4}$ Department of Astronomical Science, School of Physical Sciences, The Graduate University for Advanced Studies (SOKENDAI), 2-21-1 Osawa, Mitaka, Tokyo 181-8588, Japan\\ 
$^{5}$ Division of Theoretical Astronomy, National Astronomical Observatory of Japan, 2-21-1 Osawa, Mitaka, Tokyo 181-8588, Japan}

\date{\today}

\begin{abstract}
Both high- and low-velocity outflows are occasionally observed around a protostar by molecular line emission. 
The high-velocity component is called ``Extremely High-Velocity (EHV) flow,'' while the low-velocity component is simply referred as ``(molecular) outflow.'' 
This study reports a newly found EHV flow and outflow around MMS $5$ in the Orion Molecular Cloud 3 observed with ALMA.
In the observation, CO $J$=2--1 emission traces both the EHV flow ($|v_{\rm{LSR}} - v_{\rm{sys}}|$ $\simeq$ 50--100 $\rm{km\ s^{-1}}$) and outflow ($|v_{\rm{LSR}} - v_{\rm{sys}}|$ $\simeq$ 10--50 $\rm{km\ s^{-1}}$).
On the other hand, SiO $J$=5--4 emission only traces the EHV flow.
The EHV flow is collimated and located at the root of the V-shaped outflow.
The CO outflow extends up to $\sim$ 14,000\,AU with a position angle (P.A.) of $\sim79$\,$^\circ$ and the CO redshifted EHV flow extends to $\sim$11,000 AU with P.A. $\sim96$\,$^\circ$.
The EHV flow  is smaller than the outflow, and the dynamical timescale of the EHV flow is shorter than that of the outflow by a factor of $\sim 3$.
The flow driving mechanism is discussed based on the size, time scale, axis difference between the EHV flow and outflow, and the periodicity of the knots.
Our results are consistent with the nested wind scenario, although the jet entrainment scenario could not completely be ruled out.
\end{abstract}
\keywords{ISM: individual (OMC-3, MMS 5) --jets and outflows stars: formation --jets --outflows }

\section{INTRODUCTION}
Molecular outflows are ubiquitously observed in the early evolutionary stage of star formation.
Thus, they are regarded as key to understand the relation between mass ejection and accretion  in the star formation process.
Observations imply that there are varieties of bipolar protostellar outflows such as low- or high-velocity flows and wide-angle or collimated flows.
In addition, the most energetic flow is called as the extremely high velocity outflows (hereafter referred to as $\lq\lq$EHV flows$"$; Bachiller et al. 1990a; 1990b; 1991b; 1996.)
The EHV flows have been observed in a limited number of Class 0 and I objects, which have a short lifetime of $t\ \lesssim 10^5$\ yr (Andr\'e et al. 1993).
The EHV flows, which are normally observed in the CO emission, have a velocity of $50-200\ \rm{km s^{-1}}$ with respect to the systemic velocity and exhibit a collimated structure (opening angle of $\sim 5-20^\circ$; Gueth et al. 1996; Lebron et al. 2006; Santiago-Garcia et al. 2009; Podio et al. 2015).
The EHV outflow is also occasionally detected in the SiO emission.
All the EHV flows with SiO emission are associated with the Class 0 sources ($t\ \lesssim 10^4$ yr).
Such sources are extremely rare, and only seven samples are currently known (Bachiller et al. 1900a; 1990b; 1991a; 1996, 2001; Lebron et al. 2006; Hirano et al. 2010; Gomez-Ruiz 2013; Tafalla et al 2016; Lee et al. 2017).
The EHV flow often traces a collimated jet-like structure as well as knots within the structure (Bachillier \& Tafalla 1999).
In contrast, low velocity molecular outflows also exhibit bipolarity, but have larger opening angles of $30-60^\circ$ (Tafalla et al. 2004, 2010, 2016; Santiago-Garcia et al. 2009; Hirano et al. 2010).
Thus, the EHV flow appears to be enclosed by the low velocity outflow.

Two scenarios are proposed to explain the driving of both low and EHV flows: (i) nested disk wind (Tomisaka et al. 2002; \citealt{banerjee06}; Machida et al. 2008) and (ii) entrainment scenarios  (Arce et al. 2007 and references therein).
In the former  scenario, low- and high-velocity flows are directly driven by the inner and outer regions of the circumstellar disk, respectively (Machida et al. 2014).
On the other hand, in the latter scenario, only the high-velocity flow is accelerated near the protostar and entrains the surrounding gas until the entrained gas reaches supersonic speed and creates the low-velocity outflow (Arce et al. 2007).
In order to disentangle the two scenarios observationally, we should confirm the age and size differences between low- and high-velocity flows in the early protostellar stage.
In other words, if the former scenario is correct, we would find age differences between low-velocity (older) and high-velocity (younger) flows when the protostar is very young. 
This is because the low-velocity flow appears before the emergence of the high-velocity flow \citep{tomisaka02}. 
On the other hand, in the latter  scenario, the entrainment of outflow occurs as a consequence of the propagation of high-velocity flow.
Thus, the entrainment scenario implies that both the low- and high-velocity components have approximately the same dynamical age, even in a very early stage of star formation.

In order to specify the driving mechanism of protostellar outflows, we choose a unique EHV flow associated with MMS 5, which is  located in the Orion Molecular Cloud 3 (OMC-3), as our sample distance of $d \sim 388$ pc (Kounkel et al. 2017).
The MMS 5 (Chini et al. 1997) is also called CSO 9 (Lis et al. 1998) or HOPS 88 (Megeath et al. 2012; Furulan et al. 2016), and identified as a Class 0 source.
This object has an envelope mass of $8-36$ $M_\odot$ and a bolometric luminosity of 16 $L_\odot$ (Chini et al. 1997; Takahashi et al. 2008; Megeath et al. 2012; Furlan et al. 2016).
A compact east-west bipolar outflow (P.A.$ \sim -90^{\circ}$) is also observed in the CO $J$=1--0 and $J$=3--2 emissions, in which the outflow momentum flux is estimated to $\sim 10^{-5} M_\odot\, \rm{km\ s^{-1}\, yr^{-1}}$ (Aso et al. 2000; Williams et al. 2003; Takahashi et al. 2008).
The size of the outflow is as large as $\sim 0.1$ pc (one side).
Only low-velocity components were confirmed in the past observation (Aso et al. 2000; Williams et al. 2003; Takahashi et al. 2008).
The dynamical timescale of the low-velocity outflow was estimated to be $\sim 9,000$ yr (Takahashi et al. 2008).
Although only the low velocity component was found in previous studies, a recent APEX paper also reported the EHV flow from MMS 5 (Gomez-Ruiz et al., submitted).
The chain of knots observed in the near infrared $\rm{H_2}$ ($v$=1--0) emissions was also reported (Yu et al. 2000; and Stanke et al. 2002).
The $\rm{H_2}$ knots are distributed within 0.05 pc from the central protostar. 

As described in Sections 3.3 and 3.4, the size of the EHV flow (not low-velocity outflow) associated with MMS 5, which is reported in this paper, is 7,000 AU (0.035\,pc). 
The sizes of the EHV flows reported for other objects are in the range of 0.025--0.125\ pc (Bachiller et al. 1991a, 2001; Shang et al. 2006; Lebron et al. 2006; Hirano et al. 2010; Gomez-Ruiz 2013; Tafalla et al 2016; Lee et al. 2017).
Thus, the EHV flow associated with MMS 5 appears to be one of the most compact EHV flows reported previously.
Therefore, the MMS 5 is the best object to clarify the driving mechanisms of respective velocity flows immediately after protostar formation.
Note that, in this paper, we often refer to the EHV flow as ``the jet,''  ``the high-velocity flow'' or ``the high-velocity component'' in contrast to ``the outflow,'' ``the low-velocity flow'' or ``the low-velocity component.''

The observation method and data reduction are described in Section 2. 
We present the results obtained from 1.3 mm continuum emission and molecular line observations (dense gas and outflow tracers) in Section 3.
The driving mechanisms and properties of the observed EHV flow are discussed in Section 4.
Finally, we summarize our results in Section 5.

\section{ALMA OBSERVATIONS AND DATA REDUCTION}
\begin{longrotatetable}
\begin{deluxetable}{lccc}
\tabletypesize{\scriptsize}
\tablecaption{ALMA Observing Parameters}
\tablewidth{0pt}
\tablehead{
\colhead{Parameters} & \colhead{ACA 7-m array} & \colhead{12-m array LAR image \tablenotemark{a}} & \colhead{12-m array HAR image \tablenotemark{b}}}
\startdata
Observation date (YYYY-MM-DD) & 2016-06-30, 2016-07-12 and 2016-07-19 & 2016-06-29 & 2016-09-19 \\
Number of antennas & 10 & 48 & 40 \\
Phase center (J2000.0) & \multicolumn{3}{c}{$05^{h}35^{m}22^{s}.464$, $-05^{\circ}01'14''.304$} \\
Primary beam size (arcsec) & 46 & 27 & 27 \\
Continuum representative frequency (GHz) & \multicolumn{3}{c}{224} \\
Continuum effective bandwidth (GHz) & \multicolumn{3}{c}{1.1} \\
CO $J$=2--1 and SiO $J$=5--4 imaging velocity resolution ($\rm{km\ s^{-1}}$) \tablenotemark{c}& \multicolumn{3}{c}{5.0} \\
$\rm{C^{18}O}$ $J$=2--1 and $\rm{N_2D^{+}}$ $J$=4--3 imaging velocity resolution ($\rm{km\ s^{-1}}$) \tablenotemark{c}& \multicolumn{3}{c}{0.1} \\
Projected baseline coverage (m)& 7--47 & 10--310 & 10--3130 \\
Maximum recoverable size (arcsec) & 46 & 32 & 32 \\
On-source time (minutes) & 16 & 4 & 8 \\       
System temperature (K) & 50 $-$ 165 & 80 $-$ 220 & 60 $-$ 220 \\       
Flux calibrator & J0522-3627 & J0522-3627 & J0510+1800 \\
Gain calibrator & J0542-0913, J0607-0834 & J0542-0541 & J0607-0834 \\
Bandpass calibrator & J0538-4405, J0522-3627 & J0522-3627 & J0510+1800 \\       
\enddata
\footnotesize{ $^a$ LAR = low angular resolution (ALMA compact configuration).}
\footnotesize{ $^b$ HAR = high angular resolution (ALMA extended configuration).}\\
\footnotesize{ $^c$ Values listed in the table are imaging velocity resolution. The original velocity resolution used for the observations are mentioned in Secntion 2.}\\
\label{tab:t1-2}
\end{deluxetable}
\end{longrotatetable}
\begin{longrotatetable}
\begin{deluxetable}{lcccc}
\tabletypesize{\scriptsize}
\tablecaption{Summary of the Image Parameters for Each Figure}
\tablewidth{0pt}
\tablehead{
\colhead{Data set} & \colhead{Configuration} & \colhead{Synthesized beam size [arcsec, deg]} & \colhead{Noise level} & \colhead{Figure reference}}
\startdata
1.3mm continuum & ACA+12-m array LAR image & $1.8 \times 1.0$, -77 & 3.3 [$\rm{mJy\ beam^{-1}}$] & 1(a) \\
1.3mm continuum & 12-m array HAR image & $0.20 \times 0.16$, -22 & 0.9 [$\rm{mJy\ beam^{-1}}$] & 1(b) \\
$\rm{C^{18}O}$ $J$=2--1 & ACA+12-m array LAR image & $1.9 \times 1.1$, -76 & 46 [$\rm{mJy\ beam^{-1}} \cdot \rm{km\ s^{-1}}$] & 2 \\
$\rm{C^{18}O}$ $J$=2--1 & ACA+12-m array LAR image & $1.9 \times 1.1$, -76 & 34 [$\rm{mJy\ beam^{-1}} \cdot \rm{km\ s^{-1}}$] & 3 \\
$\rm{N_{2}D^{+}}$ $J$=4--3 & ACA+12-m array LAR image & $1.9 \times 1.1$, -76 & 27 [$\rm{mJy\ beam^{-1}} \cdot \rm{km\ s^{-1}}$]& 2 \\
CO $J$=2--1 & ACA+12-m array LAR image & $1.9 \times 1.1$, -76 & 32 [$\rm{mJy\ beam^{-1}} \cdot \rm{km\ s^{-1}}$]& 5 \\
CO $J$=2--1 & ACA+12-m array LAR image & $1.9 \times 1.1$, -76 & 730 [$\rm{mJy\ beam^{-1}} \cdot \rm{km\ s^{-1}}$]& 7(a) \\
CO $J$=2--1 & 12-m array HAR image & $0.18 \times 0.15$, -15 & 96 [$\rm{mJy\ beam^{-1}} \cdot \rm{km\ s^{-1}}$]& 7(b) \\
CO $J$=2--1 & ACA+12-m array LAR image & $1.9 \times 1.1$, -76 & 7 [$\rm{mJy\ beam^{-1}} \cdot \rm{km\ s^{-1}}$]& 8(a) \\
CO $J$=2--1 & 12-m array HAR image & $0.18 \times 0.15$, -15 & 5.7 [$\rm{mJy\ beam^{-1}} \cdot \rm{km\ s^{-1}}$]& 8(b), 9(a) \\
SiO $J$=5--4 & 12-m array HAR image & $0.21 \times 0.16$, -24 & 18 [$\rm{mJy\ beam^{-1}} \cdot \rm{km\ s^{-1}}$] & 6 \\ 
SiO $J$=5--4 & 12-m array HAR image & $0.21 \times 0.16$, -24 & 100 [$\rm{mJy\ beam^{-1}} \cdot \rm{km\ s^{-1}}$] & 7(a), 7(c)\\ 
SiO $J$=5--4 & 12-m array HAR image & $0.21 \times 0.16$, -24 & 3.7 [$\rm{mJy\ beam^{-1}} \cdot \rm{km\ s^{-1}}$] & 8(c), 9(a), 9(b)\\ 
\enddata
\label{tab:t2}
\end{deluxetable}
\end{longrotatetable}

  Our ALMA 12-m array and ACA 7-m array (Morita-array) observations were performed between 2016 January 29 and September 18 toward MMS 5 (R.A. = $05^{h}35^{m}22^{s}.464$, Dec = $-05^{\circ}01'14''.304$) with 7 separate executions.
Details of the observational parameters are summarized in Table 1.
Four molecular lines CO ($J$=2--1; 230.538 GHz), $\rm{N_2D^+}$ ($J$=3--2; 231.322 GHz), SiO ($J$=5--4; 217.105 GHz) and $\rm{C^{18}O}$ ($J$=2--1; 219.560 GHz), and the 1.3\ mm continuum emission were obtained simultaneously.
The total on source time was 16 minutes for the ACA 7-m array, 4 minutes for the ALMA 12-m array compact configuration (12-m array LAR image), and 8 minutes for the ALMA 12-m array extended configuration (12-m array HAR image), respectively.
Frequency Division Mode was used in the observation.
The ALMA correlator was configured to provide four independent spectral windows.
Spectral windows allocated for CO $J$=2--1 and SiO $J$=5--4 have a bandwidth of 468.75 MHz, while those for $\rm{C^{18}O}$ $J$=2--1 and $\rm{N_2D^+}$ $J$=3--2 have a bandwidth of 58.594 MHz, resulting in the spectral resolution of 0.282 MHz and 35.278 kHz, respectively.
This gave the velocity resolutions of 0.367 $\rm{km\ s^{-1}}$ for CO $J$=2--1, 0.390 $\rm{km\ s^{-1}}$ for SiO $J$=5--4, 0.046 $\rm{km\ s^{-1}}$ for $\rm{C^{18}O}$ $J$=2--1, and 0.048 $\rm{km\ s^{-1}}$ for $\rm{N_2D^+}$ $J$=3--2.
The channels that have no detection of the line emission are used to produce the continuum image.
After subtraction of line emissions, the effective bandwidth for the continuum emissions is approximately 1.1 GHz.

Calibration of the raw visibility data was performed by the ALMA observatory with the standard calibration method using the Common Astronomy Software Application (CASA; McMullin et al. 2007).
The calibrated visibility data were CLEANed to create the continuum and molecular line images.
For both the continuum and molecular line imagings, robust weighting with the Briggs parameter = 0.5 was used. 
Data sets obtained with the three different array configurations are combined in order to produce final images.
The achieved angular resolutions and noise levels for those presented in each figure, made with different combinations of data sets, are summarized in Table\ \ref{tab:t2}.

\section{RESULTS}
\noindent
\subsection{1.3\ mm Continuum Emission}
\begin{figure}
\centering
\includegraphics[width=\textwidth]{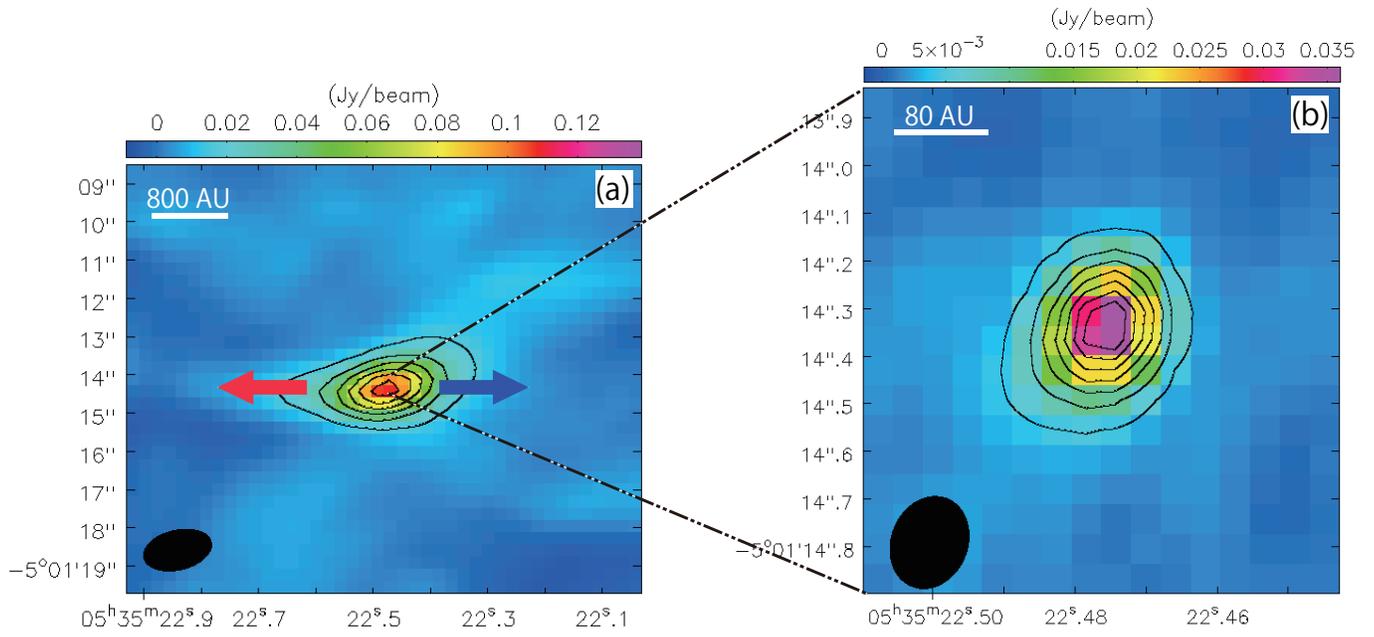}
\caption{(a) The 1.3\ mm continuum emission (color and black contours) obtained from the ACA 7-m array and 12-m array compact configuration.
The contour levels of the 1.3\ mm continuum emission start at $5 \sigma$ with an interval of $5 \sigma$ ($1\sigma$ = 3.3 mJy $\rm{beam^{-1}}$).
The red and blue arrows indicate axis of the SiO jet.
(b) The 1.3\ mm continuum emission (color and contours) obtained from the 12-m array extended configuration. The contour levels start at $5\sigma$ with an interval of $5 \sigma$ (1$\sigma$ = 0.9 mJy $\rm{beam^{-1}}$).
The ellipses in the bottom left corner show the synthesized beam size from each observation.
The spatial scale is indicated by the thick white line in the upper left corner of each panel.
}
\label{fig:1}
\end{figure}

Figure\ \ref{fig:1} presents 1.3\ mm continuum images obtained from multi-angular resolution.
The low resolution 1.3\ mm continuum image presented in Figure\ \ref{fig:1}(a) shows that the 1.3\ mm continuum emission is elongated in the east-west direction.
This elongation is consistent with the axis of the EHV flow (see Sections\ 3.3 and 3.4). 
The 1.3\ mm continuum emission enhanced along the jet axis is likely due to the hot dust associated with the jet ejected from the central region.
The total flux of the 1.3\ mm continuum emission in the low resolution image in Figure\ \ref{fig:1} (a) is 154 mJy, including the contribution from the free-free jet contained.
Assuming that the spectral index of the free-free jet is $\sim 0.6$ (Anglada et al. 1998; Reynolds 1986), the expected flux density of the free-free emission at 1.3\ mm becomes 0.94\ mJy, based on the 3.6\ cm flux density of 0.12\ mJy with 3$\sigma$ upper limit (Reipurth et al. 1999).
Hence, the flux density attributed to the free-free emission is 0.6 \% of the 1.3\ mm total flux.
The peak position is measured as R.A.\ =\ $5^h35^m22^s.464$, Dec\ =\ $-5^{\circ}01'14''.304$.
This position is offset by 0.12 arcsec with respect to the location of HOPS 88, which is identified as a Class 0 source by observations with $Hershel$ and $Spitzer$ space telescopes (Furlan et al. 2016).
The positional offset is comparable to the positional uncertainty of the $Hershel$ observation.
Thus, we consider that this continuum source is likely associated with HOPS 88.
In this paper, we refer to the position of the identified 1.3\ mm continuum source as the location of the protostar.

In the high resolution image presented in Figure\ \ref{fig:1}(b), we can confirm a compact structure associated with HOPS 88.
This compact component contains $\sim 9.5$ \% of the total flux measured from the low resolution image in Figure\ \ref{fig:1}(a).
In order to characterize the morphology of compact component, two-dimensional Gaussian fitting tool in CASA (task $\lq\lq$IMFIT") was used.
The total flux, peak flux, and deconvolved size are measured to be $57 \pm 3$ mJy, $37 \pm 1$ $\rm{mJy\ beam^{-1}}$, and $(0.14 \pm 0.02)\ \rm{arcsec} \times (0.12 \pm 0.02)\ \rm{arcsec}$ (P.A.=144 $\pm 43^{\circ}$) that corresponds to $\sim 56\ \times \sim 45$ AU in the linear size scale, respectively.
The residual after extraction of the two-dimensional Gaussian fit is less than 10 \%  as compared to the peak flux.
The mass of the circumstellar material traced by the 1.3\ mm continuum emission ($M_{1.3\rm{mm}}$) can be estimated as
\begin{equation}
M_{1.3\rm{mm}} = \frac{F_{1.3\rm{mm}}d^2}{\kappa_{1.3\rm{mm}}B_{1.3\rm{mm}}(T_{\rm{dust}})},
\end{equation}
where $F_{1.3\rm{mm}}$ is the total integrated 1.3\ mm flux of the two-dimensional Gaussian fit, $d$ is the distance to the source, $\kappa_{1.3\rm{mm}}$ is the dust mass opacity at $\lambda =$ 1.3\ mm, $T_{\rm{dust}}$ is the dust temperature, and $B_{1.3\rm{mm}}(T_{\rm{dust}})$ is the Planck function at a temperature of $T_{\rm{dust}}$. 
We adopt $\kappa_{1.3\rm{mm}} = 0.009\ \rm{cm^2\ g^{-1}}$ from the dust coagulation model of the MRN (Mathis et al. 1977) distribution with thin ice mantles at a number density of $10^6\ \rm{cm^{-3}}$ computed by Ossenkopf \& Henning (1994).
We assume a gas-to-dust mass ratio of 100.
Here, $T_{\rm{dust}}$ = 21.5 K is adopted (Sadavoy et al. 2016).
Given the measured total flux of 57 $\pm 3$ mJy from the 2D Gaussian fitting, the mass is estimated to be $M_{1.3\rm{mm}} \sim$ 1.8 $\times\ 10^{-2} M_{\odot}$.

\subsection{$\rm{C^{18}O}$\ $J$=2--1 and $\rm{N_2D^+}$\ $J$=3--2 Emission}
\begin{figure}
\centering
\begin{minipage}[c]{1\textwidth}
\includegraphics[width=\textwidth]{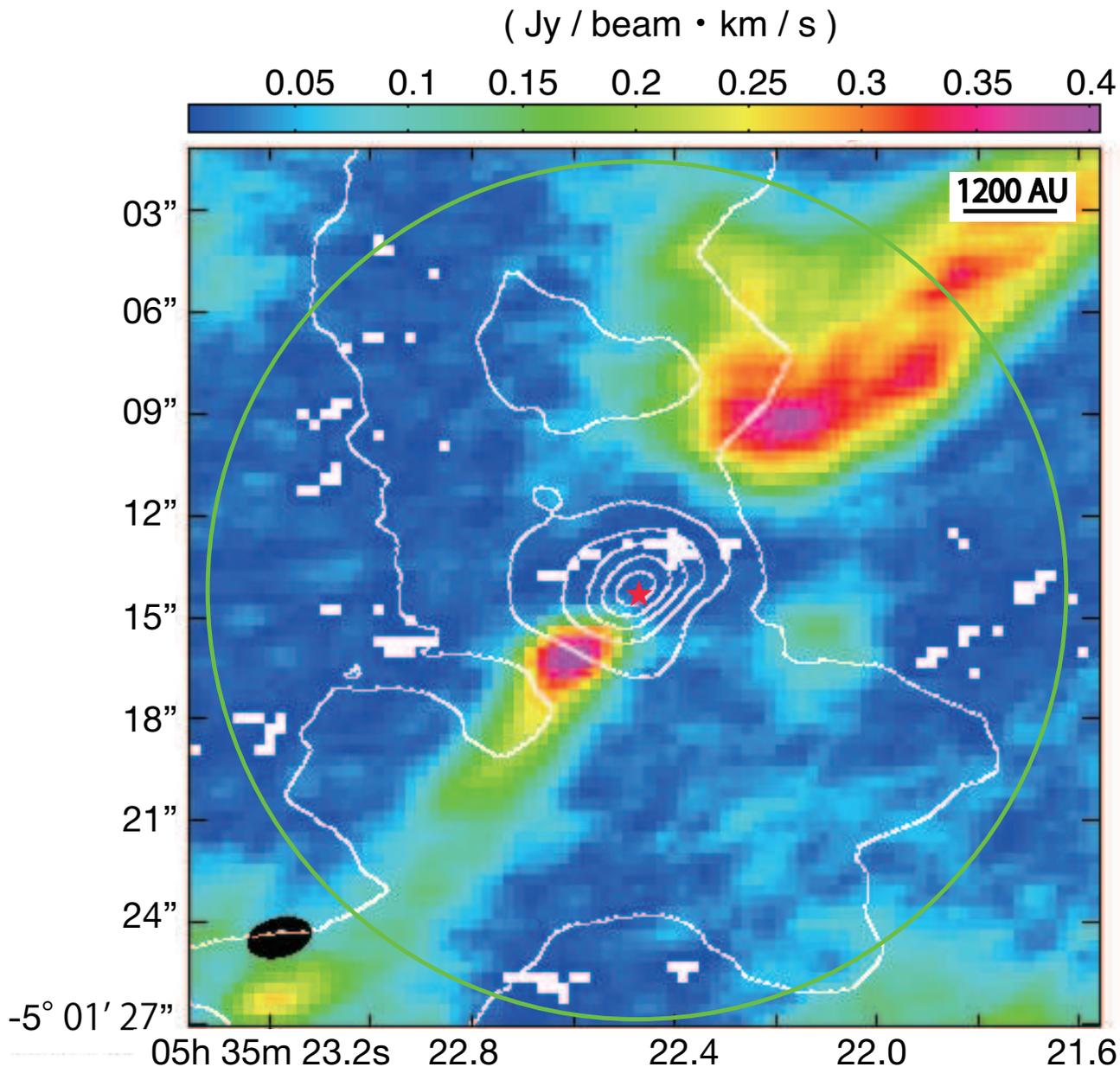}
\caption{Integrated intensity maps of $\rm{C^{18}O}$ $J$=2--1 and $\rm{N_2D^+}$ $J$=3--2 obtained from the ACA 7-m array and 12-m array compact configuration.
The white contour indicates the $\rm{C^{18}O}$ $J$=2--1, and the color indicates the $\rm{N_2D^+}$ $J$=3--2.
The contour levels start at $5 \sigma$ with an interval of $5 \sigma$ (1$\sigma$ = 46 mJy $\rm{beam^{-1}}  \cdot \rm{km\ s^{-1}}$).
The red star indicates the peak position of the continuum emission, and the green circle indicates the primary beam size (FWHM).
The black ellipse in the bottom left corner indicates the synthesized beam size of $\rm{C^{18}O}$ and $\rm{N_2D^+}$. 
The spatial scale is indicated by the thick white line in the bottom right corner.
}
\label{fig:1-2}
\end{minipage}
\end{figure}

Figure\ \ref{fig:1-2} shows the dense gas distribution traced by $\rm{C^{18}O}$\ $J$=2--1 and $\rm{N_2D^+}$\ $J$=3--2.
The $\rm{C^{18}O}$\ $J$=2--1 traces a centrally condensed compact structure associated with the protostar.
The deconvolved size and P.A. are measured to be (4.7 $\pm 0.4$) arcsec $\times (3.8 \pm 0.3$) arcsec and 100 $\pm 25^{\circ}$ based on the 2D Gaussian fitting.

$\rm{N_2D^+}$\ $J$=3--2 emission shows a large scale structure extending along the northwest-southeast direction.
Note that the gas distribution is consistent with the elongation of the OMC-3 filament (e.g., Chini et al. 1997; Lis et al. 1998; Johnstone et al. 1999). 
In contrast to the $\rm{C^{18}O}$\ $J$=2--1 emission, $\rm{N_2D^+}$\ $J$=3--2 emission is very weak in the vicinity of the protostar and no emission with 5 $\sigma$ level is detected toward the 1.3\ mm continuum emission and the $\rm{C^{18}O}$ emission peak.
The $\rm{N_2D^+}$ molecule is known to be abundant only in the cold and dense environments where molecules such as CO are frozen-out onto the surface of the dust grains forming icy mantles (Fontani et al. 2012; Giannetti et al. 2014).
The CO molecule is frozen-out onto grain mantles when the gas temperature is below $T$ $\leq 19$ K (Qi et al. 2011).
Thus, after a protostar emerges, CO appears in the gas phase due to heating by radiation from the central protostar. 
The CO in the gas phase significantly decreases $\rm{N_2H^+}$ formation and accelerates $\rm{N_2H^+}$ destruction (Bergin et al. 2001).
The anti-correlation of gas-phase CO and $\rm{N_2H^+}$ has been confirmed by numerous observations of the protostellar environment (Caselli et al. 1999; Bergin et al. 2002; J\o rgensen 2004).
Since MMS 5 harbors a protostar, it is natural to expect warm gas around the protostar, which explains the anti-correlation between $\rm{N_2D^+}$ and $\rm{C^{18}O}$ gases.
The systemic velocity measured from the optically thin $\rm{N_2D^+}$\ $J$=3--2 emission line is 11.0 $\rm{km\ s^{-1}}$, which is consistent with the central velocity of the dip in the $\rm{C^{18}O}$\ $J$=2--1 line profile.
This appears to indicate that both gases have an identical center of mass and that $\rm{N_2D^+}$\ $J$=3--2 encloses $\rm{C^{18}O}$\ $J$=2--1 around MMS 5.

\begin{figure}
\centering
\begin{minipage}[c]{1\textwidth}
\includegraphics[width=\textwidth]{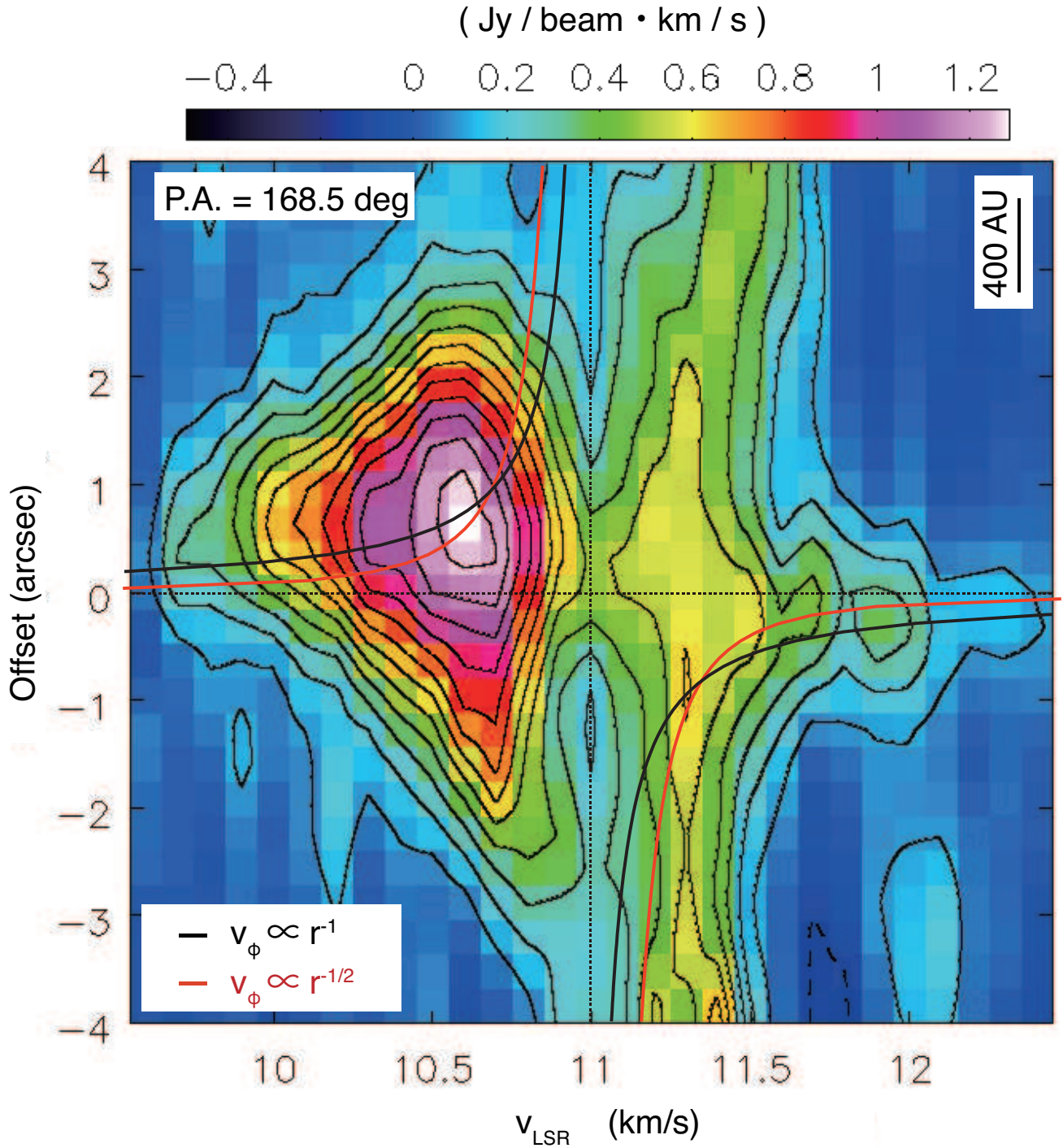}
\caption{$\rm{C^{18}O}$ $J$=2--1 PV diagrams cutting along P.A.=$169^{\circ}$ from the continuum peak position using the ACA 7-m array and 12-m array compact configurations.
The contour levels start at $3 \sigma$ with an interval of $3 \sigma$ ($1\sigma$ = 34 $\rm{mJy\ beam^{-1}} \cdot \rm{km\ s^{-1}}$).
The vertical thin dotted line indicates the systematic velocity, $v_{\rm{sys}} = 11.0\ \rm{km \ s^{-1}}$.
The horizon line indicates the peak position of the high resolution continuum image.
The curves indicate $v_{\phi} \propto r^{-1}$ (black) and $v_{\phi} \propto r^{-1/2}$ (red), respectively.
}
\label{fig:2}
\end{minipage}
\end{figure}

Figure\ \ref{fig:2} presents the position-velocity diagram cutting along P.A. = $169^{\circ}$, which is the direction perpendicular to the outflow axis. 
Figure\ \ref{fig:2} shows that the $\rm{C^{18}O}$\ $J$=2--1 spreads to  $\pm 1.3\ \rm{km\ s^{-1}}$ with respect to $v_{\rm sys}$.
The blueshifted emission is twice as brighter than the redshifted emission.
There exist intensity peaks in both the blue- and redshift components with the positional offset of $\sim 1$ arcsec with respect to the position of the 1.3\ mm continuum peak, indicating the velocity gradient along the major axis of the envelope.
The feature observed in Figure\ \ref{fig:2} can be interpreted as a rotational motion within the envelope.
In Figure\ \ref{fig:2}, we overlaid two rotation curves expected from the Keplerian rotation ($v_{\phi} \propto r^{-0.5}$) and the angular momentum conservation ($v_{\phi} \propto r^{-1}$).
The rotation curve expected from the angular momentum conservation ($v_{\phi} \propto r^{-1}$) seems to fit well and follow the emission peaks in the PV diagram. 
In addition, a high velocity gas of $|v_{\rm{LSR}}$$-v_{\rm{sys}}|$ $\geq 1\ \rm{km\ s^{-1}}$ is detected toward the center (within $\leq 0.5$ arcsec from the center) and the compact component is particularly clear in the redshifted component.
However, the $\rm{C^{18}O}$ component in Figure\ \ref{fig:2} does not contradict the velocity curve expected from both the angular momentum conservation ($v_{\phi} \propto r^{-1}$) and Keplerian rotation ($v_{\phi} \propto r^{-0.5}$) assuming the central protostellar mass of 0.1 $M_\odot$.
High angular resolution and high sensitivity observations are required in order to determine the origin of the high velocity gas within the region of 200 AU from the protostar or that at the emission peak.

\subsection{Outflow and Jet Traced by CO $J$=2--1}
\begin{figure}
\centering
\begin{minipage}[c]{1\textwidth}
\includegraphics[width=\textwidth]{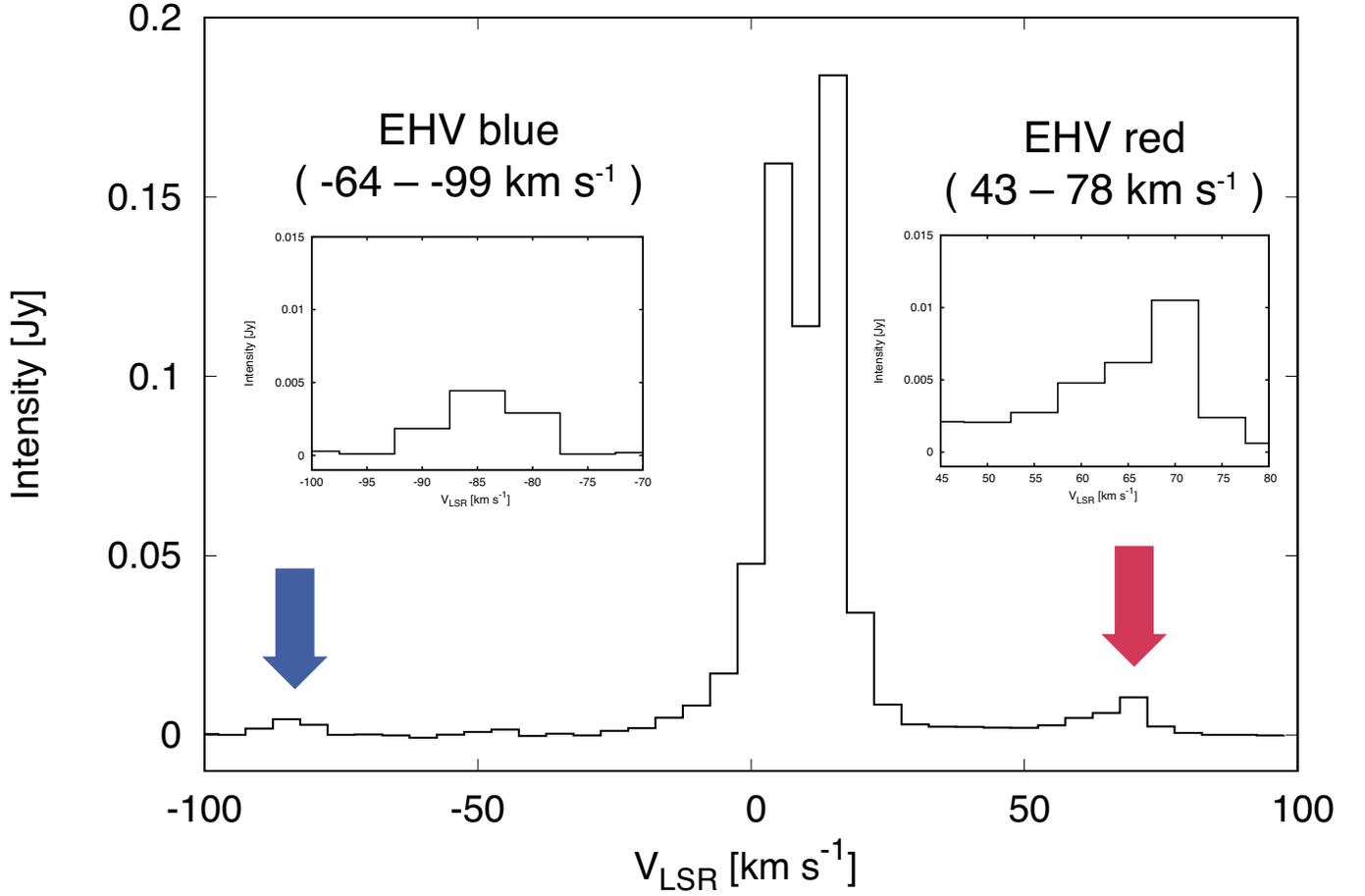}
\caption{The CO $J$=2--1 line profile, in which the main peak and two side peaks are shown.
The main peak originates from the low-velocity outflow, $|v_{\rm{LSR}} - v_{\rm{sys}}| \lesssim$ 10 -- 50 $\rm{km\ s^{-1}}$, and the two side peaks, which are indicated by arrows, originate from the EHV flow around -64 -- -99 $\rm{km\ s^{-1}}$ at the blueshifted emission and 43 -- 78 $\rm{km\ s^{-1}}$ at redshifted emission, where $v_{\rm{sys}}$ = 11.0 $\rm{km\ s^{-1}}$.
The insets indicate the line profiles of the EHV blueshifted and redshifted components.}
\label{fig:3}
\end{minipage}
\end{figure}

Figure\ \ref{fig:3} shows the CO $J$=2--1 line profile toward the peak position of MMS 5.
The figure shows low- and high-velocity components.
The main component ranges between velocities of $\pm\ 7\ \rm{km\ s^{-1}}$ with respect to the systemic velocity.
This component has a strong absorption at the center.
The velocity center of the dip is $v_{\rm{LSR}}$ = 11.0\ $\rm{km\ s^{-1}}$, which is consistent with the systemic velocity.
In addition, we have detected EHV flows in CO $J$=2--1 toward MMS 5.
The detected velocity ranges of the blueshifted and redshifted components are $-64$ -- $-99\ \rm{km\ s^{-1}}$ and 43 -- 78\ $\rm{km\ s^{-1}}$, respectively.
In other words, the EHV flow is accelerated up to the range of 80 to 100 $\rm{km\ s^{-1}}$ with respect to the systemic velocity.

The CO $J$=2--1 channel maps shown in Figure\ \ref{fig:4} indicate that the molecular outflow associated with MMS 5 is elongated along the east-west direction.
In the velocity range of $|v_{\rm{LSR}} - v_{\rm{sys}}|$ = 0 -- 10 $\rm{km\ s^{-1}}$ (Figure\ \ref{fig:4}(a)), the extended CO emission is detected rather uniformly.
The complication of the emission is attributed to contamination from the ambient molecular cloud.
In the velocity range of $|v_{\rm{LSR}} - v_{\rm{sys}}|$ = 10 -- 50 $\rm{km\ s^{-1}}$ (Figure\ \ref{fig:4}(b)--(e)), CO emission delineates a V-shaped structure.
The size of this structure extends up to $\sim 24$ arcsec, which corresponds to $\sim 0.05$ pc (P.A. = $79 \pm 2^{\circ}$) and the opening angle of the V-shape is measured to be $\sim 40^\circ$.
The outflow is elongated approximately east-west direction, velocity of $\sim 10\, \rm{km\ s^{-1}} $ and size of  $\sim 0.1$ pc, respectively, which are consistent with those estimated in  previous studies (Aso et al. 2000; Williams et al. 2003; Takahashi et al. 2008).
At $|v_{\rm{LSR}} - v_{\rm{sys}}|$ = 50 -- 100 $\rm{km\ s^{-1}}$ (Figure\ \ref{fig:4}(f)--(j)), which corresponds to the velocity of the EHV flow, a geometrically collimated structure appears.
The length and width of the collimated structure measured from the redshifted flow are $\sim 7,000$ AU and $\sim 1,200$ AU, respectively.
The synthesized beam size ($\leq$ 600 AU) is smaller than the measured outflow width, hence the collimated component is spatially resolved.
The blueshifted component also shows some degree of collimation.
The emission is not detected at the protostar position.\\

\subsection{Jet Traced by SiO $J$=5--4}
In channel maps presented in Figure\ \ref{fig:5}, we detected collimated SiO $J$=5--4 emission associated with MMS 5.
The redshifted emission is mainly detected in the velocity range of $|v_{\rm{LSR}} - v_{\rm{sys}}|$ = 50 -- 70 $\rm{km\ s^{-1}}$.
Faint SiO $J$=5--4 redshifted emission with signal to noise ratio (SNR) $\simeq$ 5 is also detected in the low velocity range of $|v_{\rm{LSR}} - v_{\rm{sys}}| \leq 50$ $\rm{km\ s^{-1}}$.
For the blueshifted emission, there is no detection (SNR less than 3) in the velocity range of $|v_{\rm{LSR}} - v_{\rm{sys}}| \leq 80$ $\rm{km\ s^{-1}}$.
In the velocity range of $|v_{\rm{LSR}} - v_{\rm{sys}}|$ = 80 -- 90 $\rm{km\ s^{-1}}$, the blueshifted EHV component is detected in Figure\ \ref{fig:5}(i).
The extension of the SiO $J$=5--4 redshifted emission is 4 arcsec, which corresponds to $\sim 1,600$ AU (P.A. = $96 \pm 1^{\circ}$) and the width along the minor axis is $\sim$ 0.7 arcsec, which corresponds to $\sim 270$ AU.
Thus, the EHV component is spatially resolved with our synthesized beam.
The comparison of the high and low angular resolution images indicates recovering $\sim 91$ \% of the total integrated flux.
Therefore, the detected SiO $J$=5--4 emission is concentrated to the collimated structure captured with the high angular resolution data.
In the western part of the outflow and jet (blueshifted component), a chain of knots are observed in near infrared ${\rm H_2}$\ $v$=1--0 line (Yu et al. 2000; Stanke et al. 2002).
The direction of the sequence of ${\rm H_2}$ knots coincides with the jet direction in our observations.
However, the chain of ${\rm H_2}$ knots are shifted to the south more than 5 arcsec compared to the location of the SiO and CO jets in our observation. 
In addition, the width  of  the ${\rm H_2}$ knots ($\sim$ 15 arcsec) closest to the MMS 5 (i.e., location of the protostar) is much wider than that  of the CO and SiO jets ($\sim$ 5 arcsec) in our observation.
These differences clearly mean that the origin of the ${\rm H_2}$ knots is  not related to the CO and SiO jets associated with MMS 5.

\subsection{Morphology Comparisons between SiO and CO emissions}
In Figure\ \ref{fig:6}(a), we compared the spatial distribution of the integrated intensity SiO $J$=5--4 emission with that of CO $J$=2--1 emission.
The SiO collimated structure (redshifted emission) is observed at the root of a V-shaped structure detected in CO $J$=2--1 emission.
The SiO $J$=5--4 collimated structure, which comes only from the EHV velocity component (80 -- 90\,$\rm{km\ s^{-1}}$), is more compact than the V-shaped CO $J$=2--1 structure by a factor of 7.5, in which the CO $J$=2--1 emission is obtained from the relatively low velocity regime ($|v_{\rm{LSR}} - v_{\rm{sys}}|$ = 10 -- 50 $\rm{km\ s^{-1}}$).
The collimated high velocity structure detected in CO (11,000 AU) is larger than that detected in SiO (2,500 AU), but still compact than the V-shaped outflow (14,000 AU).
It is natural to consider that the SiO EHV component and CO low-velocity component trace physically different flows.
In order to distinguish between the two detected components, hereafter, we use the terms $\lq\lq$jet" and $\lq\lq$outflow."
The jet is defined as a geometrically collimated emission and has the EHV range of ejection speed of $\gtrsim 50-100\ \rm{km\ s^{-1}}$ (Kwan \& Tademaru 1988; Hirth et al 1994a,b).
In contrast, the outflow is defined as a bipolar structure that has a wide opening angle and has a relatively low velocity of $\lesssim 50 \rm{km\ s^{-1}}$ (Lada 1985).

The V-shaped CO outflow has P.A.$=79^{\circ}$, while the CO and SiO jets have P.A.$\sim 90^{\circ}$.
Thus, the position angle of the jet and outflow is not the same as seen in Figure\ \ref{fig:6}(a).
The difference in position angle between the jet and outflow is about $18^{\circ}$ measured in redshifted side (for detailed discussion, see Section 4.1).
As seen in Figure\ \ref{fig:6} (b) and (c), the jet component is detected in both CO $J$=2--1 and SiO $J$=5--4 emissions.
Within both CO and SiO jets, several clumpy structures are detected.
They appear more or less periodically with a spacing of $\simeq$ 200 AU.
The interpretation of the clumpy structures is discussed in Section 4.2.

\begin{figure}
\centering
\begin{minipage}[c]{1\textwidth}
\includegraphics[width=\textwidth]{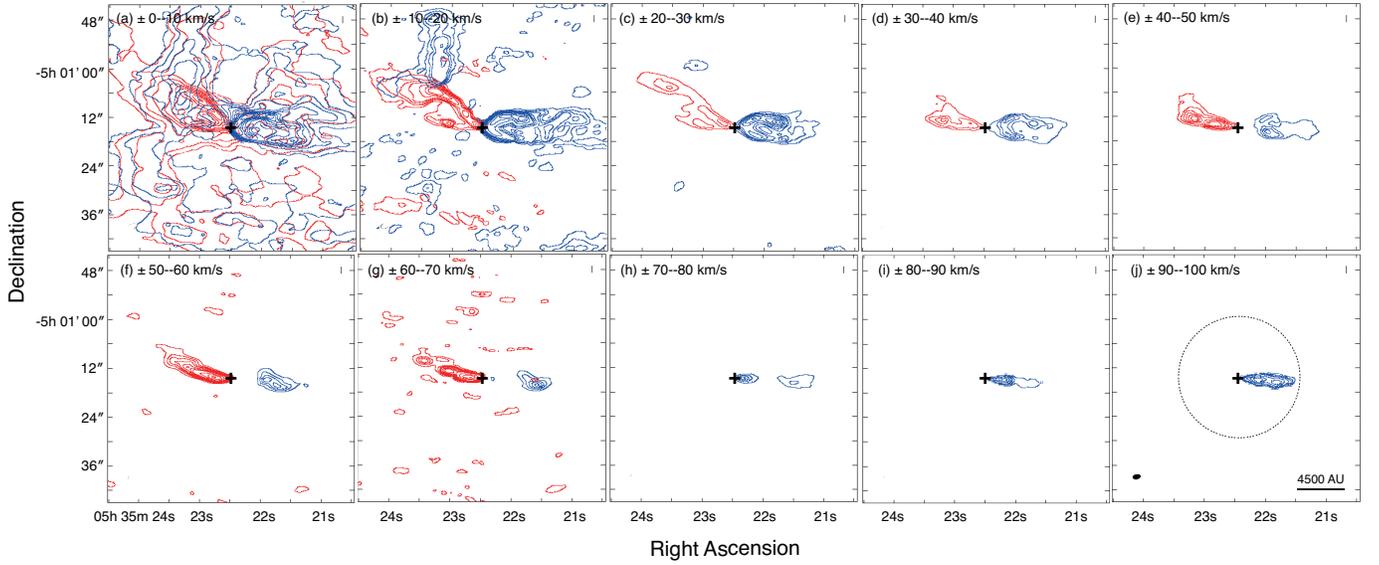}
\caption{Channel maps of the CO $J$=2--1 emission averaged over 10 $\rm{km\ s^{-1}}$ velocity intervals.
The velocity range with respect to the systemic velocity of ${v_{\rm{LSR}}} = 11.0\ \rm{km\ s^{-1}}$ is shown at the upper left corner of each panel.
The crosses indicate the 1.3\ mm continuum peak position of MMS 5.
The contour levels are (a) [$10, 50, 100, 150, 200, 250, 300, 350, 400, 450, 500 \sigma$], (b) [$10, 20, 30, 40, 60, 80, 100, 150, 200, 250 \sigma$], (c) [$10, 20, 30, 40, 60, 80, 100, 120 \sigma$], (d) [$10, 20, 30, 40 \sigma$], (e) [$10, 20, 30, 40, 50, 60 \sigma$], (f) [$10, 20, 30, 40, 60, 80, 100, 150, 200, 250 \sigma$], (g) [$10, 20, 30, 40, 60, 80, 100, 200, 300, 400 \sigma$], (h) [$10, 20, 30, 40 \sigma$], (i) [$10, 20, 30, 40, 60, 80, 100 \sigma$], and (j) [$10, 20, 30, 40, 60, 80, 100 \sigma$], respectively.
For the CO line images, 1$\sigma$ is equal to 32 mJy $\rm{beam^{-1}}$.
The circle in panel (j) shows the primary beam size (FWHM) of $\sim 26$ arcsec and the synthesized beam size (a filled ellipse) is denoted at the bottom left corner.
}
\label{fig:4}
\end{minipage}
\end{figure}

\begin{figure}
\centering
\begin{minipage}[c]{1\textwidth}
\includegraphics[width=\textwidth]{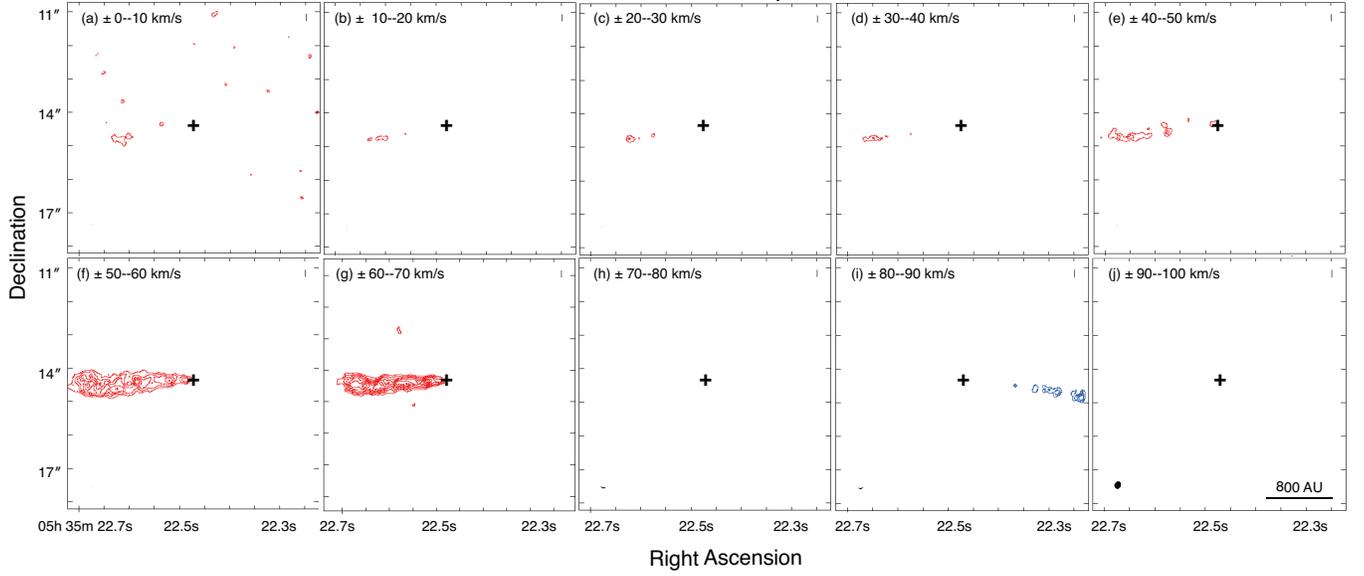}
\caption{Channel maps of the SiO $J$ = 5--4 emission averaged over 10.0\,$\rm{km\ s^{-1}}$ velocity intervals.
The velocity range with respect to the systemic velocity of ${v_{\rm{LSR}}} = 11.0\,\rm{km\ s^{-1}}$ is shown at the upper left corner of each panel.
The crosses indicate the 1.3\ mm continuum peak position of MMS 5.
The contour level begins from $5 \sigma$ with an interval of $2 \sigma$ ($1\sigma$ for SiO line images is equal to 18 mJy $\rm{beam^{-1}}$).
The synthesized beam size is denoted at the bottom left corner in panel (j).}
\label{fig:5}
\end{minipage}
\end{figure}

\begin{figure}
\centering
\begin{minipage}[c]{1\textwidth}
\includegraphics[width=\textwidth]{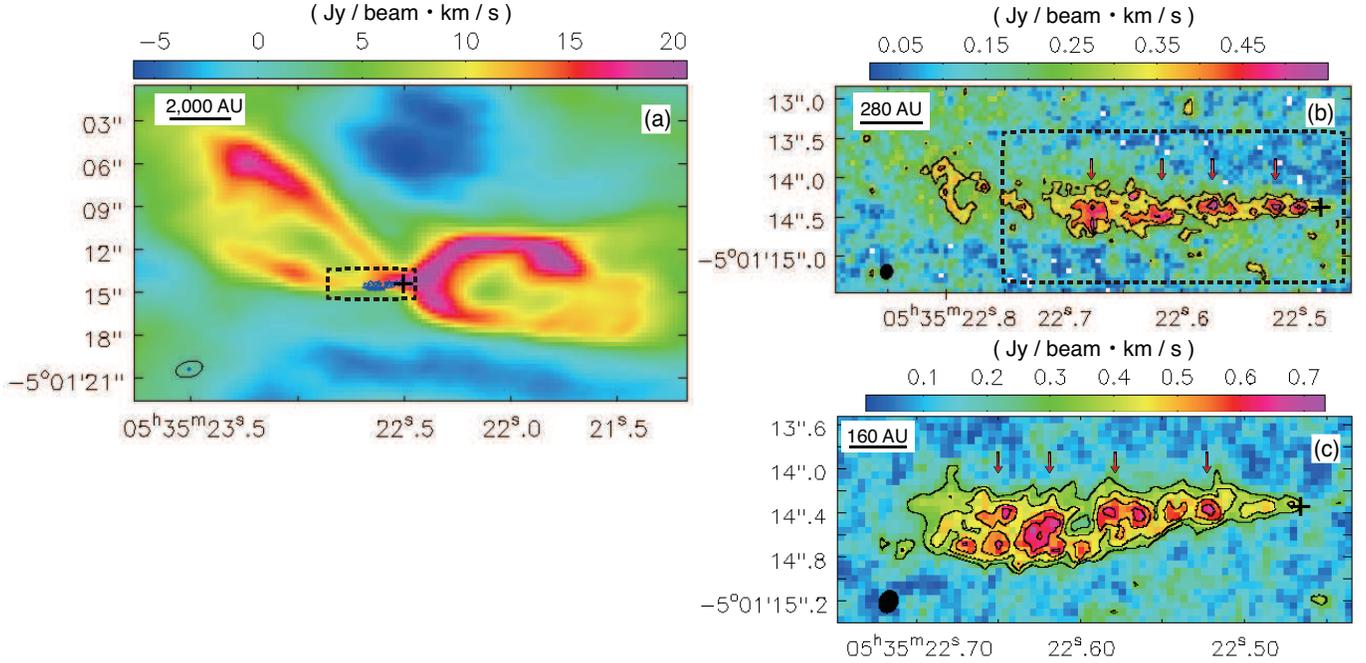}
\caption{(a) The integrated intensity of CO $J$=2--1 line (color) overlaid with that of the SiO $J$=5--4 line (blue contours) obtained from the ACA 7-m array and the 12-m array compact configuration and the 12-m array extended configuration.
The contour levels of the SiO $J$=5--4 line start at $4 \sigma$ with an interval of $1 \sigma$ ($1\sigma$ for the SiO $J$=5--4 line images is 0.1 Jy $\rm{beam^{-1}}\ km\ s^{-1}$). 
(b) Zoomed image of the redshifted component of CO $J$=2--1 using the velocity range between 50 and 80 $\rm{km\ s^{-1}}$.
The plot is made with the 12-m array extended configuration data sets (high angular resolution image).
The contour level starts at $3 \sigma$ with $1 \sigma$ interval ($1\sigma$ for the high velocity component of CO $J$=2--1 line images is equal to 0.96 Jy $\rm{beam^{-1}\ km\ s^{-1}}$).
(c) Zoomed image of the redshifted component obtained from SiO $J$=5–-4.
The plot is made with the 12-m array extended configuration (high-angular resolution image).
The contour level starts at $3 \sigma$ with $1 \sigma$ interval ($1\sigma$ for the SiO $J$=5--4 line image is equal to 0.1 Jy $\rm{beam^{-1}\ km\ s^{-1}}$).
The crosses indicate the 1.3\ mm continuum peak position of MMS 5.
The beam size is indicated by filled or emptied ellipse in each panel.
The zoomed regions are indicated by the dotted rectangles in panels (a) and (b).
The spatial scale is also shown at the upper left corner of each panel.
Knots are presented by the arrows in panels (b) and (c).
}
\label{fig:6}
\end{minipage}
\end{figure}

Figures\ \ref{fig:4} and \ref{fig:5} show asymmetry in the outflow and in the jet.
For example, in the red-lobe side, the northern emission is stronger than the southern emission.
As for the jet, the red component is stronger than the blue component.
We can not figure out the mechanism to explain these asymmetric structures only from the present observational data.
However, we suggest some possibilities for the asymmetry; a non-uniform density distribution of the surrounding medium caused by turbulence, and an initial asymmetric structure of the star-forming core. 
In the latter part of this paper, we focus on the properties of the jet and outflow, while we do not discuss the asymmetry. 

\clearpage

\clearpage

\clearpage

\clearpage

\clearpage

\clearpage

\clearpage

\clearpage
\begin{figure}
\centering
\begin{minipage}[c]{1\textwidth}
\includegraphics[width=\textwidth]{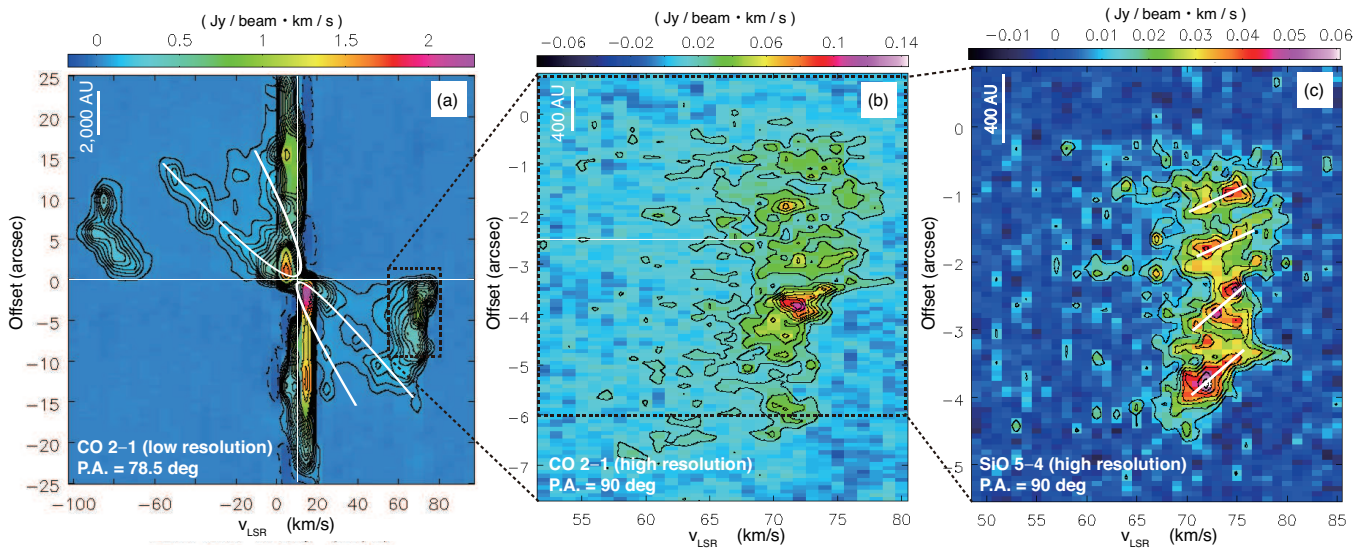}
\caption{(a) PV diagram cutting along the outflow axis, P.A.=$79^{\circ}$, obtained from CO $J$=2--1 using the ACA 7-m array and 12-m array compact configuration (low angular resolution image).
Contour levels are -5, 5, 10, 15, 20, 25, 30, 40, 50, 60, 70, 80, 90, 100, 150, and 200$\sigma$, and $1\sigma$ for CO line images is 7 mJy $\rm{beam^{-1}}$.
The white line is the fitting curve of the outflow shell structure.
(b) Zoomed image of the black dotted rectangle in panel (a) with the velocity range between 50 and 80 $\rm{km\ s^{-1}}$, but PV diagram is cutting along the jet axis, P.A.=$90^{\circ}$.
The plot is made with the 12-m array extended configuration data (high-angular resolution image).
The contour level starts at $3 \sigma$ with an interval of $2 \sigma$ ($1\sigma$ for the high velocity component of CO line images is 5.7 mJy $\rm{beam^{-1}}$).
(c) Zoomed image of the black dotted rectangle in panel (b) with the velocity range of 50 and 85 $\rm{km\ s^{-1}}$.
The plot is made with the 12-m array extended configuration (high angular resolution image) obtained from SiO $J$=5--4.
The contour level starts at $3 \sigma$ with an interval of $2 \sigma$ ($1\sigma$ for the SiO line images is 3.7 mJy $\rm{beam^{-1}}$).
The white line indicates the trend of each knot direction.
The spatial scale is shown at the upper left corner of each panel.
}
\label{fig:7}
\end{minipage}
\end{figure}

\subsection{Physical Parameters of Outflow and Jet}

\begin{deluxetable}{lccccc}
\tabletypesize{\footnotesize}
\tablecolumns{2}
\tablecaption{Outflow and Jet Parameters}
\tablewidth{0pt}
\tablehead{
\colhead{} & \colhead{$L_{\rm obs}$\tablenotemark{a} [AU]} & \colhead{$L$\tablenotemark{b} [AU]} & \colhead{$v_{\rm obs}$\tablenotemark{c} [${\rm km\,s^{-1}}$]} & \colhead{$v$\tablenotemark{d} [${\rm km\,s^{-1}}$]} & \colhead{$t_{\rm dyn}$\tablenotemark{e} [yr]}}
\startdata
outflow (CO $J$=2--1)\ldots\ldots & 9,300         & 14,000 & 30            & 40   & 1,300 \\
jet (CO $J$=2--1)\ldots\ldots\ldots      & 7,000         & 11,000 & 70            & 90  & 470 \\
jet (SiO $J$=5--4)\ldots\ldots\ldots      & 1,600         & 2,500 & 70            & 90  & 110 \\
\enddata
\footnotesize{ $^a$ Typical apparent size scale.}
\footnotesize{ $^b$ Intrinsic size scale assuming $i=50^{\circ}$.}\\
\footnotesize{ $^c$ Typical observed line-of-sight velocity.}
\footnotesize{ $^d$ Intrinsic speed assuming $i=50^{\circ}$.}
\footnotesize{ $^e$ Dynamical timescale.}
\label{tab:t3}
\end{deluxetable}

Figure\ \ref{fig:7}(a) shows the PV diagram cutting along the major axis of the outflow (P.A. $= 79^{\circ}$).
In this diagram, we found three components.
The first exists in the high velocity range between $-110$ and $-70\ \rm{km\ s^{-1}}$ for the blueshifted component and between 50 and 70 $\rm{km\ s^{-1}}$ for the redshifted component with respect to the systemic velocity.
The spatial extension of this component is about 18 arcsec ($\sim 7,000$ AU) with respect to the protostar.
This  component corresponds to the jet.
The second component shows a parabolic structure in the PV diagram (as delineated by the white line in Figure\ \ref{fig:7}(a)), having the velocity range $|v_{\rm{LSR}} - v_{\rm{sys}}|\ \leq 60\ \rm{km\ s^{-1}}$.
Although the spatial extension of this component is almost the same as that in the jet, this component corresponds to the V-shaped outflow.
The parabolic structure seems to interact with the jet around the terminal velocity ($v_{\rm jet} \sim 60 \rm{km\ s^{-1}}$) in the redshifted component.
The third component consists of very low-velocity gas ($|v_{\rm{LSR}} - v_{\rm{sys}}|\ \leq 10\ \rm{km\ s^{-1}}$), and has a larger spatial extensions than the first and second components.
The third component shows the same velocity gradient as in the outflow, which might be related to the swept-up gas by the outflow.
We will not discuss further about this component in this paper.

In order to more precisely derive the inclination angle of the low-velocity outflow with a relatively wide opening angle detected in CO $J$=2--1, we use a  simplified analytical model  (wind-driven model) introduced by Lee et al. (2000).
In cylindrical coordinates, the structure of the outflow shell is described as
\begin{equation}
z = CR^2,
\end{equation}
where $z$ and $R$ are the height from the disk mid-plane and the distance from the outflow (or $z$-) axis, respectively.
The outflow velocities in the radial ($v_R$) and $z$-direction ($v_z$) are given by
\begin{equation}
v_R = v_0\, R,
\end{equation}
and
\begin{equation}
v_z =v_0\, z,
\end{equation}
respectively.
In equations (2) -- (4), $C$ and $v_0$ are free parameters representing the spatial and velocity distributions of the outflow shell.
We also introduce the inclination angle, $i$, of the outflow shell (see Fig.21 of Lee et al. 2000).
The parameter $C$ is determined by the spatial structure of the emission from the outflow.
After $C$ is determined, the parameters $v_0$ and $i$ can be estimated to consider the inclination effect in the PV diagrams, in the same manner as in the previous study (Lee et al. 2000) \footnote{The angle $i$ represents the angle between the outflow axis and the celestial plane.}.

Hereafter, we use the following observational datasets to derive the physical parameters of the outflow/jet.
For the CO emission, we use both blueshifted and redshifted emissions to estimate the physical parameters and drive their mean values.
Since SiO emission is only detected in the redshifted component, the physical parameters of SiO jet is calculated only using the one-side component.
The curvature parameter is fit to the outflow shells as $C$ = 0.07 $\rm{arcsec^{-1}}$.
Then, based on the measured $C$, the outflow structures give $v_0 = 6.7\ \rm{km\ s^{-1}\ arcsec^{-1}}$, and the inclination angle of the outflow axis with respect to the plane of the sky was $i=50^{\circ}$.

Using the derived inclination angle of the outflow shell, we estimated the sizes, velocities, and dynamical time scales of the outflow and jet which are listed in Table 3. 
The dynamical timescale is estimated by the intrinsic length scale $L$ and the expansion speed $v$ as $t_{\rm dyn} \sim L/v \sim (L_{\rm obs}/v_{\rm obs})\ \rm{tan}\, 50^{\circ}$, where $L_{\rm obs}$ is the projected length of the outflow/jet and $v_{\rm obs}$ is the line-of-sight velocity of the outflow/jet.
The projected maximum size of CO $J$=2--1 and SiO $J$=5--4 emission, $L_{\rm obs}$, was measured at the 10 $\sigma$ contour in their channel maps (Figures\ \ref{fig:4} and \ref{fig:5}).
The maximum redshifted gas velocity measured in Figure\ \ref{fig:7}  were used as $v_{\rm jet,obs}$ of the jet, while the mean velocity of CO outflow ($\sim 30\ \rm{km\ s^{-1}}$) measured in Figure\ \ref{fig:7}(a) was used for $v_{\rm out,obs}$ of the outflow.
The outflow has a dynamical timescale of $t_{\rm dyn}\ \sim 1,300\, (L_{\rm out,obs}/9,300\, {\rm\, AU})(v_{\rm out,obs}/30\, {\rm km\, s}^{-1})^{-1}$ yr.
The jet dynamical timescale is estimated to $t_{\rm dyn} \sim 470\, (L_{\rm jet,obs}/7,000\, {\rm\, AU})(v_{\rm jet,obs}/70\, {\rm km\, s}^{-1})^{-1}$ yr, using CO $J$=2--1.
Thus, the dynamical timescale of the outflow is factor of $\sim 3$ longer than that of the jet.
In addition, the molecular outflow ($\sim 14,000$ AU) is larger than jet ($\sim 11,000$ AU).
Using the timescales and size scales, we discuss the driving mechanism of this outflow/jet system in Section 4.1.

Figure\ \ref{fig:7}(b) and (c) show zoomed images of PV diagrams of the redshifted jet component obtained in the CO and SiO emissions, respectively.
The high-velocity component located in the velocity range of $|v_{\rm{LSR}} - v_{\rm{sys}}|$ = 55 -- 70 $\rm{km\ s^{-1}}$ traces the EHV jet.
Within the EHV jet of SiO in Figure 8(c), we can see several bright compact knots.
As guided by the white lines in Figure\ \ref{fig:7}(c), each bright knot has a velocity gradient inside when cutting along the jet direction.
This is consistent with the result of the jet simulation by Stone and Norman (1993), who considered the case of the episodic jet.
They obtained a similar saw tooth velocity field, in which the line-of-sight velocity is decelerated in a bright knot while increasing the distance from the central source.
Santiago-Garc{\'{\i}}a et al. (2009) reported the similar saw tooth structure in their observations.
Our result presented in Figure\ \ref{fig:7}(c) shows a similar pattern of the velocity gradient  (both the size scale and the velocity change). 
This saw tooth structure implies that the observed EHV jet in the SiO emission traces the unsteady or episodic gas ejection.


\begin{figure}
\centering
\begin{minipage}[c]{0.9\textwidth}
\includegraphics[width=\textwidth]{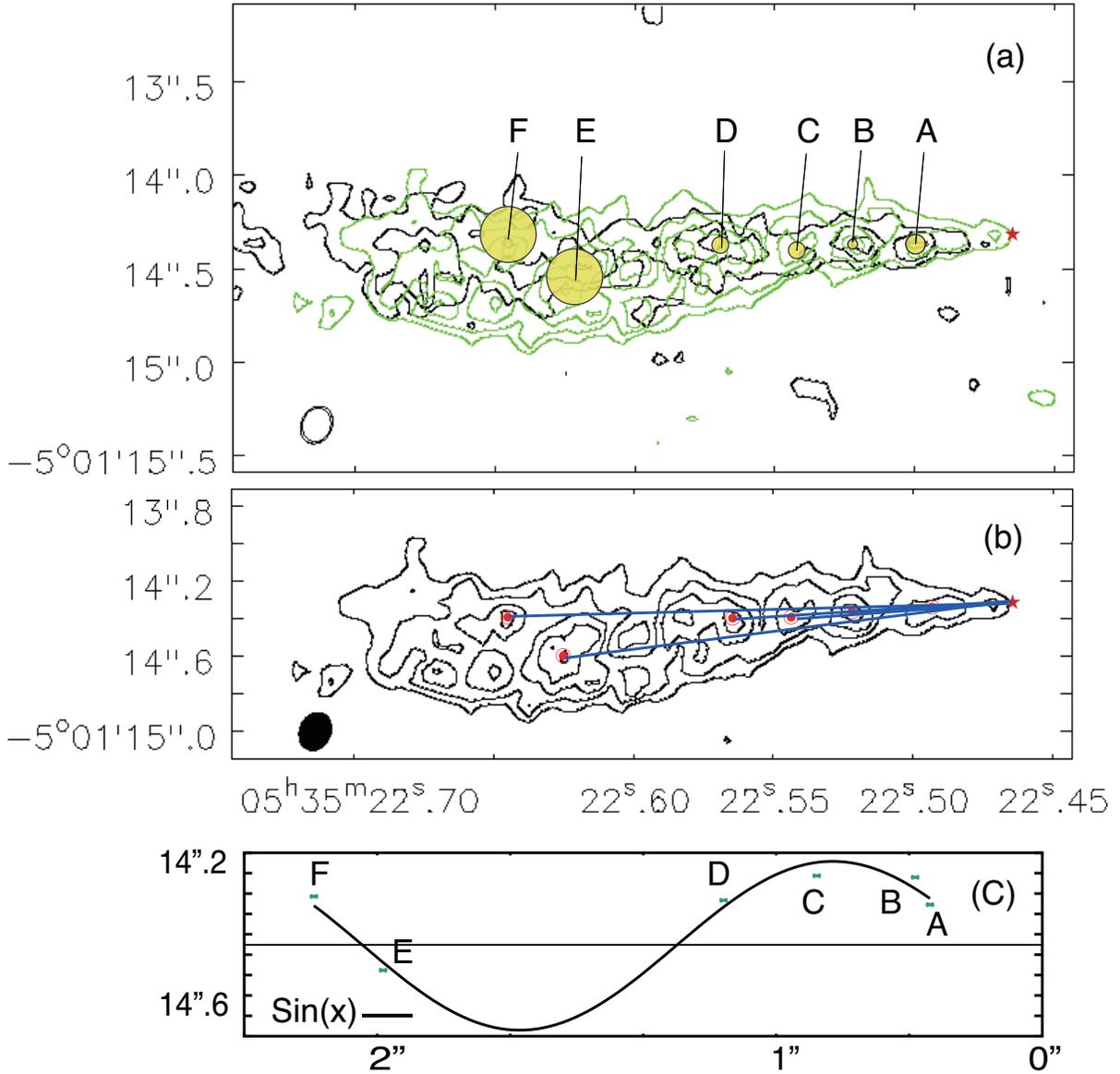}
\caption{(a) CO $J$=2--1 integrated intensity map (black) overlaid with the SiO $J$=5--4 integrated intensity map (green).
The yellow circles represent the SiO $J$=5--4 and CO $J$=2--1 peak flux positions, in which the circle size indicates a consistent level (i.e., positional error) determined by $\sim \theta_{\rm beam}/(2\,\rm{SNR})$  with SNR $\sim 3$. 
$\theta_{\rm beam}$ is the beam size.
The knots observed in CO are well coincident with those in SiO with a yellow circle. 
Between CO and SiO emissions, the positional differences of knots A, B, C, and D are $\lesssim 20$ \%, and those of knots E and F are $\sim 50$ \% of the positional error radii.
(b) The SiO $J$=5--4 integrated intensity map, in which the peak flux positions are plotted by red dots.
The blue lines connect the central continuum peak position to the SiO peak flux positions.
(c) Deviation from the jet axis for six knots, A through F, is fitted by a sine function.}
\label{fig:6-2}
\end{minipage}
\end{figure}

\section{DISCUSSION}
\subsection{Driving Mechanism of Outflow and Jet}
Two scenarios are proposed to explain the driving mechanism of the outflow and the jet, namely, (i) nested disk wind and (ii) entrainment scenarios.
In the nested disk wind scenario, different types of flow are driven from different radii of a rotationally supported disk (Tomisaka et al. 2002; \citealt{banerjee06}; Machida et al. 2008; \citealt{tomida13}) in which the low- and high-velocity flows are driven near the disk outer and inner edges, respectively (Machida et al. 2014).
On the other hand, in the entrainment scenario, only the high-velocity jet appears near the protostar and the ambient (infalling) material is accelerated or entrained by the jet (Arce et al. 2007 and references therein).
In both scenarios, the high-velocity component (i.e. jet) is considered to be driven by the Lorentz and centrifugal forces (Kudoh \& Shibata 1997; Spruit et al. 1997; Tomisaka 2002; Machida et al. 2008a; Seifried et al. 2012).
The low-velocity outflow is also magnetically accelerated by the rotation of the outer disk region in the nested disk wind scenario, while it is entrained by the jet driven near the protostar in the entrainment scenario.

In the nested disk wind scenario, since the low-velocity outflow appears before the emergence of high-velocity flow and protostar, the low-velocity outflow is expected to have a longer dynamical time than the high-velocity jet. 
In addition, the low-velocity component (i.e. the outflow) should precede the high-velocity component (i.e. the jet) even if only in the very early phase of the star formation (Machida 2014). 
However, the high-velocity jet overtakes the low-velocity outflow after a short time because the jet velocity is much higher than the outflow velocity. 
Thus, the nested disk wind scenario predicts that the jet length is shorter than the outflow length for a very short duration immediately after the protostar formation. 
Instead, in the entrainment scenario, since the jet entrains the surrounding gas and makes the low-velocity outflow, the jet length should be comparable to or larger than the outflow length at any time after protostar formation.
Thus, only the observation of low- and high-velocity components in a very early phase of the star formation could determine their driving mechanism.

In our observation, the outflow and jet images have an angular resolution of 0.2 arcsec, which corresponds to $\simeq 80$ AU.
The outflow and jet launching points are expected to be located at $\gtrsim 2$ AU and $\lesssim 0.5$ AU from the central star (\citealt{tomisaka02}; Machida 2014).
Recent ALMA studies have shown that an outflow with a wide opening angle appears in the outer-disk region, $\sim 100$ AU (Alves et al. 2017), and a jet with good collimation is driven near the disk's inner edge, $\sim 0.05$ AU (Lee et al. 2017).
The velocities of these flows are orders of 10 ${\rm km\ s^{-1}}$ and 100 ${\rm km\ s^{-1}}$, respectively.
Although we are not able to spatially resolve the outflow and jet launching points (or radii) due to the limited angular resolution, the observed gas morphologies and gas velocities share similar characteristics.
This indicates that MMS 5 outflow and jet might be launched from the different radii.

As described in Section 3.5, the dynamical timescale of the jet is shorter than that of  the outflow by factors of 3.
The outflow and the jet extend up to $\sim 14,000$ AU and $\sim 11,000$ AU, respectively.
The size of the jet is shorter than that of the outflow.
This is consistent with the nested disk wind scenario.

Another possible evidence to support the nested disk wind scenario is the axis difference between the outflow and the jet ($\delta \theta\ \sim $ $17^{\circ}$; Fig $\ref{fig:6}$).
This difference between the jet and outflow axes could be explained by considering different launching radii.
The axis difference between the outflow and the jet can be explained in recent MHD simulations (Matsumoto et al. 2017, \citealt{lewis18}), in which a warped disk forms in a weakly turbulent cloud and the direction of low-velocity outflow changes with time \citep[see also,][]{matsumoto04}.
In these studies, the outflows are not always aligned with a large-scale disk, including both Keplerian and pseudo disks.
The disk normal on the scale of $\gtrsim 100$ AU is roughly aligned with the outflow axis, while that on the scale of $\lesssim 100$ AU is misaligned with the outflow axis (Fig.9 of Matsumoto et al. 2017).
Since the protostar was not resolved and no high-velocity jet appears in Matsumoto et al. (2017), we cannot confirm difference between outflow and jet axes.
However, the inclination of the disk and thus the direction of gas flow ejected from the disk would depend on their scale.
Thus,  it is possible to expect a difference between the jet and the outflow axes in the nested disk wind scenario.

In our observation, the outflow dynamical timescale is approximately three times longer than that of the jet (see Table 3), hence the outflow is considered to be more evolved.
We also consider the possibility that a disk precesses in a short timescale of $\lesssim 10^3$\,yr.
Since the outflow traces the mass ejection history for the last $\sim 10^3$\,yr (dynamical timescale), the change of the outflow axis can be interpreted as the change of  the normal direction of the disk (i.e. precession) where the outflow is driven.
On the other hand, since the gas associated with the jet traces only the mass ejection history around the protostar in the recent $\sim 10^2$ yr, the direction of the jet indicates the instantaneous directions of the angular momentum and the poloidal magnetic field of the disk in the vicinity of the protostar, which would not be related to a long-term precessing motion observed in the long-lived outflow.


In the nested disk wind scenario, the jet appears immediately after protostar formation.
Thus, the protostellar age tends to be the same as the dynamical time of the jet.
However, when the jet age is very young, there is also the possibility that we only detected the recently driven jets and missed the pre-existing jets.
Although we cannot deny the possibility of the disappearance of preexisting long-lived jets, we could not find any signature of the high-velocity component (jet) larger than the low-velocity component (outflow). 
In summary, although the jet entrainment scenario could not completely be ruled out, there are useful evidence to support the nested disk wind scenario such as the difference of the size, dynamical time scale and axis between the outflow and the jet. 
Our observation implies that the outflow and jet are driven by different radius as expected in the nested disk wind scenario.

\subsection{Episodic Mass Ejection of Jet}
As shown in Figure~\ref{fig:6}, we found several knots in the high-velocity jet (or EHV flow). 
These types of knots are confirmed in other EHV flows observed by molecular line emission (Santiago-Garcia et al. 2009; Kwon et al. 2015). 
Usually, these knots are explained by episodic mass ejection from the region near the protostar (e.g. Stone \& Norman 1998).  
The CO jet observed in MMS 5 is comparable to the smallest known EHV flows (e.g. $\sim 5,000$ AU in Hirano et al. 2010) and the SiO jet is smaller than these flows by a factor of 3 -- 5.
Thus, our target is one of the youngest among known EHV flows.

\begin{deluxetable}{lcccccc}
\tablecaption{Jet Knots Parameter}
\tablewidth{0pt}
\tablehead{
\colhead{knots parameter} & \colhead{A} & \colhead{B} & \colhead{C} & \colhead{D} & \colhead{E} & \colhead{F}}
\startdata
    distance [arcsec] & 0.48 & 0.85 & 1.21 & 1.50 & 2.50 & 2.74 \\
    spacing [arcsec] & 0.48 & 0.37 & 0.35 & 0.29 & 1.0 & 0.25 \\
\enddata
\label{tab:t4}
\end{deluxetable}

In order to analyze these knots, we superimposed the SiO integrated intensity map on the CO integrated intensity map in Figure\ \ref{fig:6-2}(a).
We identified six knots from both SiO and CO emissions and estimated the spacing between neighbor knots, as listed in Table 4.
The average and median spacings are measured to be $\Delta \theta \simeq$ 0.46 and 0.36 arcsec, which correspond to $\Delta L\, =\, \Delta \theta \cdot d\, / \cos\, i\, \simeq$ 280 and 220 AU, respectively, with $i$ = $50^{\circ}$.
The intrinsic jet velocity is estimated to be $v$ = $v_{\rm jet,obs}/\sin\, i\, \simeq\, 70\, {\rm km\, s^{-1}}/\sin\, 50^{\circ}\, \simeq\, 90\, {\rm km\, s^{-1}}$.
Considering that the knots are caused by the episodic mass ejection, the protpstar ejects gas every $\sim$ 9 -- 12 years.

In addition, as presented in Figure~\ref{fig:6-2}(b), the knots appear to wiggle within the jet.
The wiggling is fitted by a simple sine curve in Figure~\ref{fig:6-2}(c), which may be precession of the jet or disk around the protostar.
Note that, as seen in Figure~\ref{fig:6-2}, the knots are roughly aligned along the jet axis except for ``knot E.''
Note also that although we fitted the wiggling as a simple since curve for simplicity, we need further high-resolution observations to more precisely determine the mechanism.
Assuming that the sine curve covers one cycle of the precession, the period of the cycle can be estimated to be $P \sim 50$\,yr ($\sim$ jet one cycle length / jet velocity $v_{\rm jet}$). 
Assuming the Keplerian disk around the protostar, we can estimate the typical radius inducing the precession. 
Using the Keplerian rotation period $P = 2\pi/\Omega$ and Keplerian angular velocity  $\Omega = ( GM_{\ast}/r^3 )^{1/2}$, we can estimate the precession radius as 
\begin{equation}
r_{\rm{disk}} = 6.3 \biggl( \frac{M_{\ast}}{0.1\, M_{\odot}} \biggr)^{1/3} \biggl(\frac{P}{50\, {\rm yr}} \biggr)^{-2/3}\, {\rm AU} ,
\end{equation}
to fit the rotation on period as $P \sim 50$\, yr.
When the protostellar mass of 0.1 $M_\odot$ is assumed\footnote{The outflow dynamical timescale is $\sim 10^3$ yr, and the mass accretion rate $\sim 10^{-4} M_\odot\ \rm{{yr}^{-1}}$ is assumed.}, the precession radius becomes approximately 6\,AU. 
Now, we cannot specify the cause of the existence of the precession motion at the radius $r_{\rm disk}$.
However, there are some possibilities or expectations.
The radius $r_{\rm disk}$ may correspond to the size of the rotationally supported disk.
Alternatively, the gravitational instability or magnetic dissipation may be induced at $r_{\rm disk}$ (Machida 2014). 
Moreover, a binary companion may exist around $r_{\rm disk}$.
We need further high-resolution observations to identify the cause of the precession.

Finally, we estimated the jet launching radius based on the jet velocity. 
We assume that the jet velocity approximately corresponds to the Keplerian rotation velocity at the launching radius (Kudoh \& Shibata 1997),
\begin{equation}
{v_{\rm{Kep}}}^2 = \frac{GM_{\ast}}{r_{\rm{jet}}}.
\label{eq:jetvel}
\end{equation}
From equation~(\ref{eq:jetvel}), assuming $v_{\rm{Kep}}$ is equal to the intrinsic jet velocity $v_{\rm{jet}}$, the launching radius is estimated as 
\begin{equation}
r_{\rm{jet}} = 1 \biggl( \frac{M_{\ast}}{0.1M_{\odot}} \biggr) \biggl( \frac{v_{\rm{jet, obs}}}{70\, \rm{km\ s^{-1}}} \biggr)^{-2} \left(\frac{\sin\, 50^{\circ}}{\sin\, i}\right)^2  R_\odot.
\label{eq:rjet}
\end{equation}
Although there are uncertainties in deriving equation~(\ref{eq:rjet}), such as the inclination angle and the protostellar mass, our result indicates that the jet appears near the surface of the protostar.

\section{CONCLUSION AND FUTURE PROSPECTS}
\noindent
We present ALMA observations of a unique EHV flow discovered in MMS5/OMC-3.
The main results are summarized as follows:
\begin{enumerate}
\item We detected a compact structure, which is estimated to be $1.8 \times 10^{-2}\ M_\odot$ in the 1.3\ mm continuum emission.
The structure is more or less perpendicular to the outflow and jet axes and likely traces a disk-like structure.
$\rm{C^{18}O}$ $J$=2--1 also traces a 2,000 AU scale centrally concentrated structure.
In contrast to $\rm{C^{18}O}$ $J$=2--1 emission, $\rm{N_2D^{+}}$ $J$=3--2 emission is very weak around the protostar and is elongated in the direction of the OMC-3 filament.
The PV diagram of $\rm{C^{18}O}$ $J$=2--1 shows the rotation of the disk-like structure.

\item CO $J$=2--1 emission shows two typical structures of the low-velocity outflow ($|v_{\rm{LSR}} - v_{\rm{sys}}|$ = 10 -- 50 $\rm{km\ s^{-1}}$) and the high-velocity collimated jet ($|v_{\rm{LSR}} - v_{\rm{sys}}|$ = 50 -- 100 $\rm{km\ s^{-1}}$).
The outflow and the jet extend up to $\sim 14,000$ AU and $\sim 11,000$ AU, respectively.
The size of the the jet is shorter than that of the outflow.

\item Deriving the inclination angle, we estimated the sizes, velocities, and dynamical time scales of the outflow and the jet.
The dynamical timescale of the jet is $\sim 3$ times shorter than that of estimated for the outflow.
In addition, the difference in P.A between the outflow and the jet is $17^{\circ}$.
The misalignment between the jet and the outflow can be explained if their launching radii and epochs are different.
Thus, it is natural to consider that the driving regions and driving epochs of the outflow and the jet are different.

\item Six knots are identified in both the SiO and CO emissions, which seem to have periodicity. 
The average and median values of the spacing are measured to be 0.46 and 0.36 arcsec, respectively.
Assuming the jet velocity of 90 $\rm{km\ s^{-1}}$ with the correction of the inclination angle, the jet is considered to be driven from the vicinity of the protostar every 9 -- 12 years.
In addition, as a whole, the jets have the wiggle structure. 
The wiggle can be fitted by a sine curve, implying the precession of the jet and the disk.
The precession timescale and precession radius are estimated to be $\sim 50$\,yr and $\sim 6$\ AU, respectively.
Moreover, using the jet velocity, we estimated the launching radius of the jet, which corresponds approximately to the solar radius.
\end{enumerate}
For the driving mechanism of low- and high-velocity flows, our results seem to be consistent with the nested disk wind scenario, although we cannot rule out the entrainment scenario.
The estimated launching point (or radius) is as  small as those expected in the MHD simulations (i.e., the nested wind model; Tomisaka 2002, Machida 2014).
Moreover, the jet driving radius is also more or less consistent with that estimated from the recent molecular jet observations toward HH211 (Lee et al. 2017).

\section{ACKNOWLEDGMENT}
This paper uses the following ALMA data: ADS/JAO.ALMA \#2015.1.00341.S. ALMA is a partnership among ESO (representing its member states), NSF (USA) and NINS (Japan), together with NRC (Canada), MOST and ASIAA (Taiwan), and KASI (Republic of Korea), in cooperation with the Republic of Chile. The Joint ALMA Observatory is operated by ESO, AUI/NRAO and NAOJ. 
Y. Matsushita and K. Tomisaka are very grateful for support from the Joint ALMA Observatory (Santiago, Chile) Science Visiter Program while visiting co-author S. Takahashi.
This work was supported by JSPS KAKENHI Grants (17K05387, 17H06360, 17H02869 and 15K05032).
Y. Matsushita was supported by the ALMA Japan Research Grant of NAOJ Chile Observatory, NAOJ-ALMA-192.

\end{document}